\documentclass[journal]{vgtc}              

\onlineid{0}

\vgtccategory{Research}

\title{The CAPED Framework: Characterizing User Skills in\\ \rev{Chart} Authoring
through the Lens of Tasks and Tools}

\author{%
  \authororcid{Bongshin Lee}{0000-0002-4217-627X}, 
  \authororcid{Zhicheng Liu}{0000-0002-1015-2759},
  \authororcid{John Thompson}{0000-0002-3102-4035},
  \authororcid{Chenglong Wang}{0000-0002-5933-6620}
}

\authorfooter{
  \item
  	Bongshin Lee is with the Department of Computer Science and Engineering and the Yonsei Institute of Digital Health, Yonsei University, Seoul, Republic of Korea.
  	E-mail: b.lee@yonsei.ac.kr.
  \item
  	Zhicheng Liu is with the University of Maryland, College Park, MD, USA.
  	E-mail: leozcliu@umd.edu.
  \item 
    John Thompson is with Autodesk Research, GA, USA.
  	E-mail: john.thompson@autodesk.com.
  \item
    Chenglong Wang is with Microsoft Research, WA, USA.
    E-mail: chenglong.wang@microsoft.com.
}

\abstract{In this paper, we introduce a new framework for characterizing the user skills required for \rev{chart} authoring. Despite growing attention to authoring systems, we lack a conceptual framework that characterizes the diverse and multifaceted skills involved and helps identify opportunities for system design and development. Our initial investigation revealed that skill is a complex and elusive concept, and that existing theories on skills from other disciplines are too high-level to capture the importance of tasks and tools in \rev{data} visualizations. To address this gap, we formulate the CAPED framework that characterizes skill as the ability to create a chart involving five dimensions pertinent to tasks and tools: input data, output chart, paradigm, abstraction, and expression. We demonstrate the framework's descriptive power by highlighting different skills involved in similar authoring situations and identifying the types of knowledge associated with \rev{chart} authoring. We further present examples illustrating how the framework can be used to identify new research opportunities in \rev{supporting} users with varying skills.
}


\teaser{
  \centering
  \includegraphics[width=\linewidth, alt={The CAPED framework with five dimensions: Input Data, Output Chart, Paradigm, Abstraction, and Expression.}]{{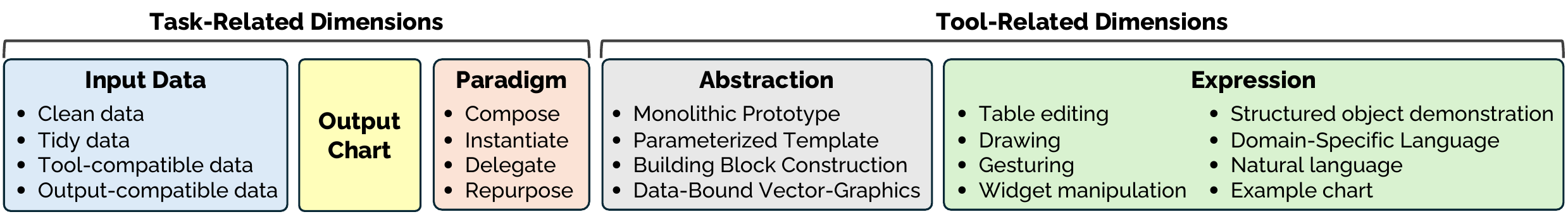}}
  \caption{The CAPED framework enables the characterization of user skills in \rev{chart} authoring using five dimensions: Input Data, Output Chart, Paradigm, Abstraction, and Expression.}
  \label{fig:teaser}
}

\graphicspath{{figs/}{figures/}{pictures/}{images/}{./}} 

\usepackage{tabu}                      
\usepackage{booktabs}                  
\usepackage{lipsum}                    
\usepackage{mwe}                       

\usepackage{mathptmx}                  

\usepackage{subcaption}
\usepackage{xspace}
\usepackage{array}
\usepackage{tabularx}
\usepackage{colortbl}
\usepackage{hhline}
\usepackage{makecell}
\usepackage{multirow}
\usepackage{graphicx}
\usepackage{url}
\usepackage[normalem]{ulem}

\newcommand{\caped}[1]{\textit{#1}}

\newcommand{\eg}{{e.g.,}\xspace}

\newcommand{\etal}{{et al.}\xspace}

\newcommand{\bpstart}[1]{\vspace{1mm} \noindent{\textbf{#1.}}}

\usepackage{soul}
\usepackage[svgnames]{xcolor}

\definecolor{RevBlue}{RGB}{9,51,220} 
\definecolor{RemRed}{RGB}{139,0,0} 

\newcommand{\rem}[1]{}
\newcommand{\rev}[1]{{#1}}

\begin{document}



\maketitle

\section{Introduction}

Creating data visualization is an essential task in both data analysis and presentation, and it has become a cornerstone of \rev{data} visualization research. 
However, compared to efforts that characterize and measure visualization literacy for consumption (e.g.,~\cite{Lee17:VLAT,Beschi25:VisLiteracyLitReview,Hong25:LLMsVisLiteracy}), the research community has given little attention to the user skills required for \rev{chart} authoring.  
Numerous toolkits, grammars, models, interactive tools, and natural language interfaces have been developed for \rev{chart} authoring, and many of them boast a reduced learning curve and lower barrier of entry. 
However, these efforts are system-centric, and it is unclear what user skills are needed to be able to use such tools and interfaces. The creators of interactive authoring tools often simply identify their target audience based on roles (e.g., designers) or implicitly characterize them using terms such as ``novices'' or ``non-experts''~\cite{burns2023we}. Such terms are ambiguous, often overloaded, and do not identify the skills the users already possess or need to acquire.

More importantly, we lack a conceptual framework to describe and discuss user skills for \rev{chart} authoring to differentiate diverse, multifaceted user profiles. For example, one user may be familiar with various types of \rev{charts}, but has no coding experience; another may be an avid programmer, but has limited knowledge in \rev{chart} design; a third may be a professional graphic designer, but rarely incorporates structured data into their work. Beyond diverse user skill sets, \rev{chart} authoring is inherently complex, involving multiple tasks (e.g., data preparation, visual encoding, \rev{chart} construction and refinement) across tools ranging from programming languages to interactive systems. As a result, different users may tackle the same authoring task using different tools, with different strategies and workflows (\Cref{sec:scope_scenario}). 


\begin{figure*}[h]
\centering
    \begin{subfigure}{0.32\textwidth}
        \includegraphics[width=\textwidth]{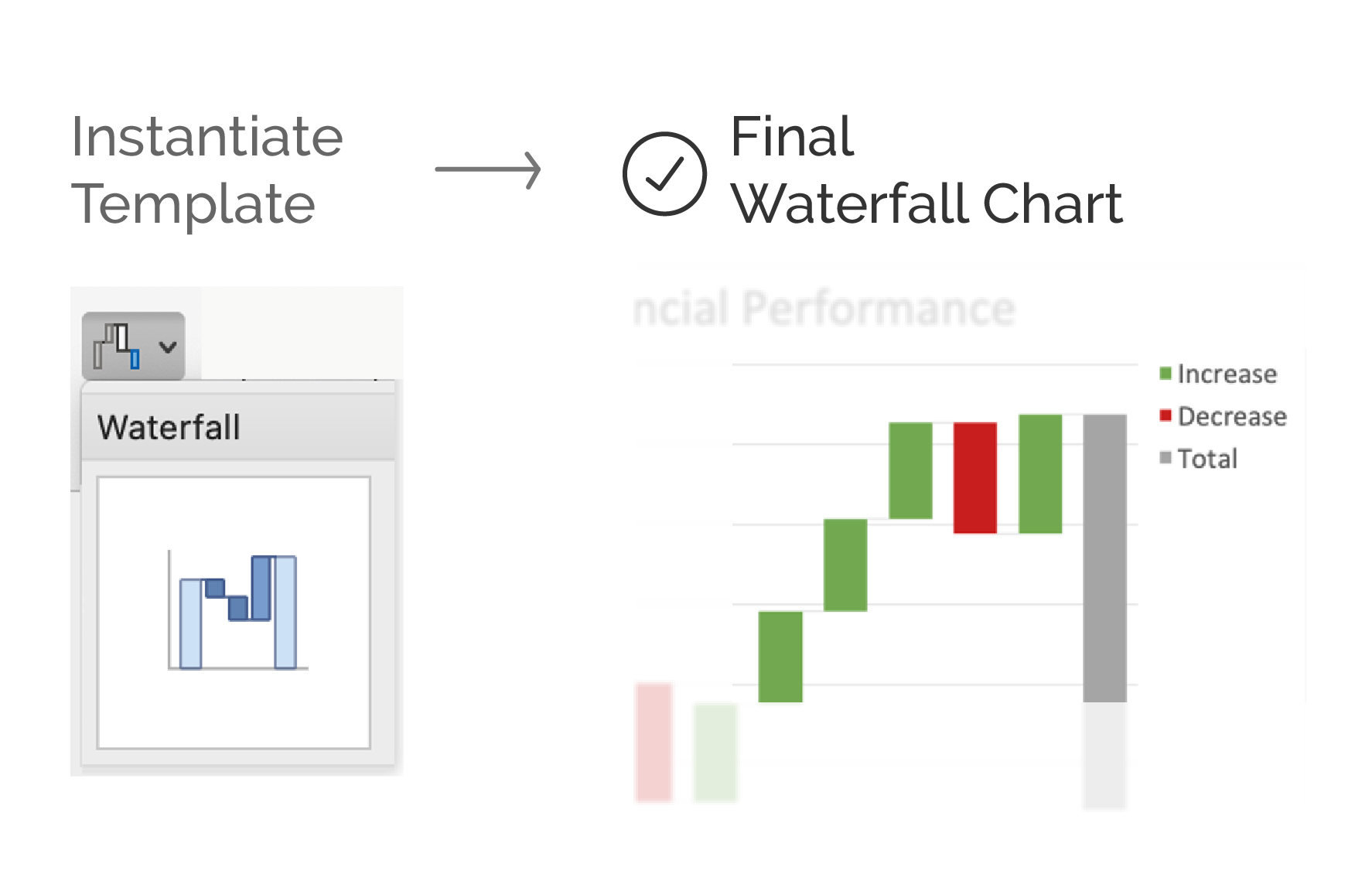}
        \caption{\label{fig:scenario_waterfall}}
    \end{subfigure}
    \hfill
    \begin{subfigure}{0.67\textwidth}
        \includegraphics[width=\textwidth]{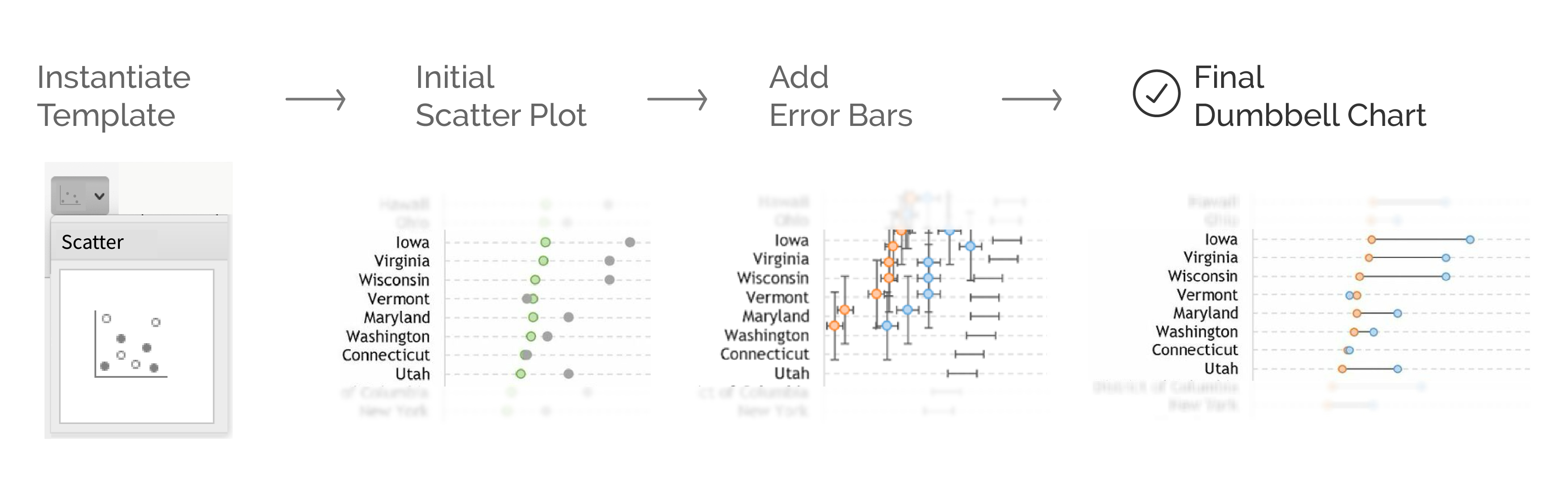}
        \caption{\label{fig:scenario_dumbbell}}
    \end{subfigure}
    \vspace{-2mm}
    \caption{An author can instantiate a template to create a waterfall chart in Excel (a), or need non-trivial efforts to ``hack'' scatter plot and error bar to compose a dumbbell chart (b).}
\end{figure*}

In this work, we aim to have a more systematic way to characterize the user skills required for \rev{chart} authoring, with the goal of helping the design and evaluation of \rev{chart} authoring systems. 
We began by \rem{synthesizing} \rev{reviewing} research on skill from sociology and learning sciences (\Cref{sec:skill_theory}), \rev{which suggests viewing skill as a conceptual construct representing the ability to accomplish a task. We then analyzed chart authoring tasks, but found that focusing on tasks alone was insufficient, as the properties of authoring tools also influence the skills required (\Cref{sec:tool_importance}). To address this, we examined high-level tool characteristics and synthesized these insights into an integrated account through the lens of authoring tasks and tools (\Cref{sec:synthesis}).}


\rem{In this paper, we present the} \rev{The resulting} CAPED framework\footnote{CAPED suggests being ``equipped'' or having a special skill, akin to being a hero with a cape, symbolizing ability and expertise. The name is composed of one letter from each of the five dimensions: output \textbf{C}hart, visualization \textbf{A}bstraction, \textbf{P}aradigm, \textbf{E}xpression, and input \textbf{D}ata.} characterizes skill as the ability to create a chart across five core dimensions---input data, output chart, paradigm, abstraction, and expression---pertinent to tasks and tools (Figure~\ref{fig:teaser}, \Cref{sec:framework}). \rev{CAPED is not a taxonomy of skills, nor a process model of chart authoring. Rather, it identifies dimensions that determine what abilities are required in authoring situations.}
We demonstrate the CAPED framework’s descriptive power by highlighting different skills involved in similar authoring situations and identifying the types of knowledge associated with \rev{chart} authoring (Section~\ref{sec:describe}). 
In addition, we use the framework to identify research opportunities \rem{leveraging AI }for \rev{supporting} different users with varying skills (Section~\ref{sec:identify}).
We conclude with the discussion on the implications and limitations of the CAPED framework, which indicate interesting future research.
By providing a unified vocabulary for authoring competencies, the CAPED framework offers a foundational step toward a more human-centered evaluation of visualization tools in an increasingly AI-driven landscape.

\section{Scope and Motivating Scenarios}
\label{sec:scope_scenario}

\subsection{What \rev{Chart} Authoring Entails in Our Framework}
\label{sec:scope}
\rev{Chart} authoring corresponds to composing a \rev{chart} with a design in mind and data in hand to portray specific data variables or insights. 
It is a necessary step in \rev{data} visualization for both communication and 
exploratory \rem{data }analysis.
Our framework characterizes the skills required in \rev{chart} authoring across both contexts\rem{, communicative and exploratory}. 
We recognize that \rev{chart} authoring is an iterative process, where authors typically brainstorm, reflect, critique, and revise a chart until it satisfies their goals.
However, we decided not to cover design ideation skills, instead focusing on the more demonstrable action of constructing a \rev{chart}.
\rev{In addition, our framework is scoped to chart authoring with abstract data in the information visualization domain.}

Authoring typically encompasses all designed aspects of the data experience, this includes the data-to-visual representations, annotations, interactions, and animations. 
However, as a first step toward developing a vocabulary to describe and discuss authoring skills, we focus on static visual representations and leave other aspects, such as interaction, annotation, and animation, as future work.
We also scope our framework to encapsulate only digital tools.
While hand-drawn~\cite{lupi_dear_2016}, painted, or fabricated visualizations are expressive mediums for representing data, this practice has significantly different backgrounds and skills compared to authoring with graphical user interfaces (GUIs).
In addition, our framework scopes data processing skills based on whether or not the end goal is to prepare the data for composing a \rev{chart}. 
We do not include statistical data analysis skills that involve methods, tools, and techniques to analyze data in the absence of \rev{charts}.



\subsection{Example Scenarios Requiring Diverse Skills}
\label{sec:scenarios}
\rev{Chart} authoring involves several different activities, including data preparation, chart construction, and chart refinement, as well as a broad range of tools from programming libraries to interactive tools. One person may have different levels of expertise for each of these activities and associated tools. In this section, we describe three example scenarios to showcase the diverse required skills in visualization authoring. \rev{We refer to these three scenarios throughout the paper as running examples. To complement the descriptions in this section, the chart images together with the full scenario text are available on our supplemental website: \url{https://caped-framework.github.io/}.}

\vspace{1mm}
\noindent\href{https://caped-framework.github.io/#scenario-01}{\textbf{Scenario 1: Two Different Charts with the Same Tool}}

\noindent Alex is an analyst who is asked to create \rev{a waterfall and dumbbell chart for a  quarterly financial performance report.} Alex decides to use Excel, a tool they are familiar with, to create these charts. \rev{A waterfall chart will illustrate changes in the company's net income per quarter. Alex prepares the data to conform to Excel's waterfall chart template: grouping by quarter and computing the difference from the previous quarter. With the data properly formatted, Alex produces the chart by selecting ``Waterfall'' chart from the ``Insert'' tab in Excel (Figure \ref{fig:scenario_waterfall}).} 

\rev{A dumbbell chart will show the company's actual performance against its targets. Excel lacks a dumbbell chart template, so Alex searches online and decides to follow a tutorial to create a dumbbell chart with work-arounds\footnote{https://simplexct.com/how-to-create-a-dumbbell-chart-in-excel} (Figure \ref{fig:scenario_dumbbell}). Following the tutorial, Alex first organizes the data to ``Insert'' a ``Scatter Plot''. As a work-around, Alex then inserts error bars to connect the target and actual points. Finally, Alex formats the chart to remove unwanted visuals from the error bars to achieve a dumbbell chart.}

\noindent\textit{\textbf{Remarks.}} Given that Alex can fluently use Excel templates to create the desired waterfall chart, a non-trivial chart, one might consider Alex is \textit{skilled} in chart authoring with Excel. However, Alex's struggle with dumbbell chart authoring contradicts this assessment. In fact, such struggle does not mean they are \textit{unskillful} with Excel either, since the two chart creation processes employ completely different authoring approaches: the first involves instantiation of built-in templates from the tool, while the latter requires ``hacking'' to \rev{appropriate} other plots to mimic the dumbbell chart. 
This scenario highlights that chart authoring skills \emph{cannot be characterized solely at the level of the tool}. Even with the same tool, different tasks may require different authoring approaches and, consequently, different user skills. \rem{When assessing skills, it is more appropriate to evaluate skills at a finer granularity: what approach is supported by a tool for a certain task and whether the user has skills with this certain approach in the tool.}

\vspace{1mm}
\noindent\href{https://caped-framework.github.io/#scenario-02}{\textbf{Scenario 2: Skills Transfer between Tools}}

\noindent\rev{Kori is a data engineer at a public health organization.  Kori is asked to create a report that explains the leading causes of death for citizens from 1970 until the present year.
The dataset includes mortality rates based on the cause of death (\eg~Cancer, Heart Disease, AIDS) for each year, age group, state residence, and gender.
Kori has a programming background and extensive experience creating data visualizations using libraries, however, Kori's organization exclusively publishes with Microsoft Power BI.}

Kori decides to visualize the change in mortality rates across age groups. They create a multi-series line chart of ``Age Group'' displaying ``Year'' on the $x$-axis and ``Mortality Rate'' on the $y$-axis. Noticing a steady yet gradual decline in all age groups, Kori wonders if the orders of magnitude of the line chart are overshadowed by the older age groups. Kori hypothesizes that a rate of change would allow better comparison. For this chart, Kori needs a new column of ``Mortality rate change (since 1970)''. 
\rev{They opt to write Python script to transform the data. Using the Pandas data library, Kori loads the dataset, computes the new column, and exports the resulting dataset. After loading the dataset (with all the required fields) back into Power BI, Kori uses drag-and-drop interactions to create a line chart that shows on the whole, all age groups exhibit similar trends of decrease, except for a noticeable peak for the ``25 to 44'' age group that coincides with the AIDS epidemic.} \rem{However, Kori now notices a peak in the ``25 to 44'' age group increasing in 1983 to 1995. After some iterations back and forth between Pandas and Power BI, Kori finds that the AIDS epidemic explains the peak. By transforming the data in Pandas, Kori learns that AIDS caused over 40,000 deaths at the peak in 1995, more than 30,000 of whom were ``25 to 44'' years old.}

\noindent\textit{\textbf{Remarks.}} Despite being relatively new to Power BI, Kori quickly adapts to it based on experience \rev{from declarative programming libraries (\eg{} Altair)}.
Chart authoring skills are transferable, beyond the authoring medium (in this case Kori was more familiar with textual code \rev{rather} than a shelf-construction UI), as Altair and Power BI share similar authoring concepts of mapping data columns to visual channels. However, transfer of skills is not bi-directional as someone with only Power BI experience would need to learn additional programming skills to use Altair. Thus, when characterizing skills, we need to consider not only data visualization concepts but also the \rev{medium} to express these concepts.
This scenario also highlights the importance of data transformation in the chart authoring process. While this step is often overlooked, it becomes essential when authors need to test hypotheses and prepare the desired charts. In Kori's case, they relied on a familiar programming paradigm to transform data. Alternatively, they could have used Power Query (a companion tool of Power BI) to perform the transforms without incurring interoperability costs.

\vspace{1mm}
\noindent\href{https://caped-framework.github.io/#scenario-03}{\textbf{Scenario 3: Different Authoring Skills to Work with AI}}

\noindent Jesse, an analyst experienced in both programming and visualization design, \rev{wants to compare four different job offers based on: Annual Salary (\$), Commute Time (minutes), Vacation Days, Remote Work, and Retirement Match.} Jesse uploads the dataset to an AI assistant and \rev{prompts}: ``Create a bar chart to compare these job offers.'' The AI assistant defaults to generating Python code for a standard grouped bar chart, directly comparing these values side by side. However, Jesse notices \rem{that the chart is not quite right:}only one bar (Annual Salary) is shown for each offer\rem{, the other variables are missing}. \rem{Wondering whether the code produced by the AI assistant is wrong or there is an issue with the data,}\rev{Jesse examines the code generated by the AI assistant, and confirms that the code is correct but still cannot see the other missing data. Realizing these five metrics span markedly different value ranges--many values are not visible to the eye--Jesse concludes a grouped bar chart is inappropriate in this context.} 

\rem{Jesse realizes that evaluating multivariate continuous data across disparate scales requires tracking individual items across multiple normalized axes.}
\rev{Jesse now provides a specific follow-up prompt to avoid} ambiguous interpretation from the AI assistant: ``Change this to a parallel coordinates plot. Give each variable its own vertical axis, normalize the scales, \rem{so we can see the relative highs and lows across each dimension,} and color the lines based on the job offer.'' The AI assistant \rem{processes this more detailed instruction,} handles the complex data normalization and plotting logic, and successfully generates a readable parallel coordinates plot. Jesse can now compare the jobs across all dimensions.

\noindent\textit{\textbf{Remarks.}} With LLMs, Jesse can generate complex data visualizations like a parallel coordinates plot. Thus, it is tempting to assume that the AI assistant ``removes'' the chart authoring skill requirement entirely. However, Jesse's success does not come from the AI agent; it relies on a combination of programming and visualization composition knowledge. While the AI accelerated the generation of complex code, Jesse's programming expertise was essential to first inspect \rev{and confirm code was syntactically correct but inappropriate in terms of chart design.} Jesse then needed the conceptual knowledge to diagnose why the grouped bar chart failed and the precise visualization vocabulary to instruct the AI on the appropriate \rem{structural} fix. 
Automated tools do not eliminate traditional skills; rather, they transform them. Chart authoring skills must evolve to encompass code verification, targeted prompting, structural critique, and the ability to steer AI agents away from conceptually flawed \rev{defaults and} toward appropriate solutions.

\section{Background and Related Work}

This section briefly reviews prior work related to chart authoring skills, including system-centric perspectives, user characterization, task taxonomies, and visualization literacy.

\bpstart{System-Centric Perspectives of \rev{Chart} Authoring Systems}
Many \rev{chart} authoring systems are developed to lower the barrier for non-technical users (or ``novices''), and they are often evaluated through user studies using traditional performance metrics such as task completion time and success rate. However, these studies typically focus on usability and do not enable systematic comparisons across authoring tools with similar goals. Furthermore, they offer limited \rev{insights into the process, required skills, and cognitive aspects of authoring tasks}~\cite{ren2018reflecting}. 
The critical reflections method~\cite{satyanarayan_critical_2019} was proposed to address some of these limitations by enabling in-depth analysis and comparison of the components and authoring processes in multiple systems. Nevertheless, it remains system-centric, is better suited for assessing a small number of authoring tools, and does not generalize to systems that embody diverse authoring paradigms and expressions. It thus falls short of supporting systematic discussions or guidelines covering diverse tools from a user’s perspective. 
Our CAPED framework provides a starting point for a user- and skill-centered characterization of \rev{chart} authoring.



\bpstart{Target User Characterization} 
\textit{Novices} have been popular target audience for data visualization: many visualization research projects often use terms such as novices and non-experts to refer to target users with little experience or skills in chart authoring~\cite{burns2023we}. For example, Elias and Bezerianos' work~\cite{elias2011exploration} reports the design and evaluation of Exploration Views, a system that allows novice visualization users to easily build and customize Business Intelligence information dashboards, and Grammel et al.'s work\cite{grammel2010information} uses the term novices to characterize people with little experience or skills.

Burns et al.~recently investigated how visualization researchers define novices and evaluate visualizations intended for them, revealing a substantial gap between the broad language used to describe novices and the narrow population representing them in evaluations (i.e., young people, students, and US residents)~\cite{burns2023we}.
The term ``non-experts'' has similar issues and lacks precision. Some \rev{chart} authoring research identifies job roles (e.g., graphic designers) when characterizing their target audience, but they still under-specify the backgrounds and skills of people in the same profession. The CAPED framework provides a vocabulary to more precisely describe user skills along the five proposed dimensions, highlighting the fact that user characterization requires more nuanced approaches.

\bpstart{Visualization Task Characterization}
Visualization task taxonomies emerged from the need to design and evaluate visualizations from a user-centered perspective. Early work focused on relating tasks to data and interaction, such as Shneiderman's ``task by data type'' taxonomy~\cite{Shneiderman96:EyesHaveIt}, Valiati et al.'s low-level tasks for multidimensional visualization~\cite{Valiati06:TaxonomyMultidimensional}, and Amar et al.'s taxonomy of low-level analytic tasks~\cite{Amar05:LowLevelComponents}. These frameworks helped describe what users can do with visualizations and supported empirical comparisons between techniques. Later work introduced higher-level views of visualization design. Tory and M\"oller framed visualizations as data transformation processes~\cite{Tory04:Rethinking}, while Munzner's nested model framed domain problems, data and task abstraction, visualization encoding, and visualization algorithms as tiered steps in the design process~\cite{Munzner09:NestedModel}. Brehmer and Munzner further bridged levels of task granularity through their multi-level typology~\cite{brehmer2013typology}, and these ideas were synthesized in Munzner's broader framework for visualization analysis and design~\cite{Munzner14:VisualizationAnalysisDesign}.

More recent work has shifted from describing visualization tasks to describing how people author visualizations~\cite{Huron14:ConstructiveVis, Mendez17:BottomUpTopDown, Wang23:NLAuthoring}. Authoring paradigms on visualization construction~\cite{Huron14:ConstructiveVis} and bottom-up versus top-down authoring~\cite{Mendez17:BottomUpTopDown} are especially relevant. Our work builds on this direction by characterizing \rev{chart} authoring in finer detail through five core dimensions (Figure~\ref{fig:teaser}, Section~\ref{sec:framework}).

\bpstart{Assessment of Literacy or Ability} Visualization literacy involves a person's ability to read, understand, and use visualizations to answer questions~\cite{Beschi25:VisLiteracyLitReview}. We contend that literacy is a prerequisite to authoring a visualization. Boy et al.~\cite{Boy14:PrincipledVisLiteracy} and Lee et al.~\cite{Lee17:VLAT} introduced the first literacy tests to quantitatively evaluate a user's comprehension abilities.
More recently, research on visualization literacy has evaluated visual encoding ability~\cite{Ge25:AVEC}, pushing the definition of literacy borrowed from textual literacy to one based on skillsets~\cite{Solen25:LiteracyOrSkillset}.
While our approach does not provide a metric for skills, it deconstructs visualizations based on data, abstractions, and paradigms. 
As we discuss in Section~\ref{sec:discussion}, we believe our framework could support the development of authoring skill assessments similar to literacy tests.
We also envision that a deeper understanding of chart authoring skills will help advance AI-assisted chart authoring. For example, Hong et al. took a crucial first step by evaluating the visualization literacy of LLMs~\cite{Hong25:LLMsVisLiteracy}. Building on our framework, we identify new research opportunities at the intersection of AI and chart authoring in Section~\ref{sec:identify}.

\section{Framework Development and Skill Framing}
\label{sec:exploration}


Motivated by the lack of a theoretical framework to describe and discuss chart authoring skills, we adopted an iterative conceptual synthesis approach involving weekly discussions among the four authors over a 16-month period.   
We began by reviewing existing theories of human skill and competence from other fields and examining their applicability to chart authoring. This review suggested construing skill as the ability to successfully accomplish a task (\Cref{sec:skill_theory}). 
Guided by this perspective, we next analyzed common activities and tasks involved in transforming input data to output charts. Through this analysis, we found that task descriptions alone were insufficient to characterize chart authoring skills (\Cref{sec:tool_importance}). We therefore conducted a comparative analysis of various chart authoring tools, identifying four recurring authoring approaches and abstracting common properties across tools. Finally, we synthesized these insights into a unified framework organized around the lens of tasks and tools (\Cref{sec:synthesis}). 

During this process, our focus emerged on viewing skill as a conceptual construct rather than a measure of proficiency (e.g., novice versus expert). 
The resulting framework is not a taxonomy of skills; rather, it identifies  dimensions that determine what skills are required in a chart authoring situation. Consequently, the dimensions in the framework should not be interpreted as skills themselves. For example, a desired output chart is not a skill; rather, a specific output chart may require knowledge and abilities such as understanding the structure of the target chart and determining whether it can be created using a given tool.
We also note that our work does not address levels of skill or expertise.


\subsection{Importance of Tasks in Skill Characterization}
\label{sec:skill_theory}

Multiple disciplines, including psychology, sociology, and economics, have tried to define and characterize the concept of ``skill'' \cite{attewell1990skill,green2011skill} or ``competence'' \cite{le2005competence}. Many view skill as the ability to accomplish a task, and the level of skill is associated with the complexity of the task: ``\textit{to exercise
greater skill is to carry out a more complex activity}'' \cite{attewell1990skill}. Similarly, competence refers to ``\textit{the ability to successfully perform a range of tasks to a high level of performance}'' \cite{green2011skill}. While economists tend to view skill and competence as interchangeable concepts, psychologists view skill ``\textit{in a narrower way to whether some can do some task or set of tasks}'' and define competence as a broader concept \cite{le2005competence}. This broader view of competence identifies four dimensions \cite{le2005competence}: \textit{cognitive} (know-that and know-why, which include both conceptual and tacit knowledge), \textit{functional} (skills or know-how, things that a person should be able to do or demonstrate), \textit{social} (know-how-to-behave, possession of appropriate values and ability to make sound judgments), and \textit{meta} (ability to acquire the previous three competencies, cope with uncertainty, and learn and reflect). 

These four dimensions bear resemblance to theoretical frameworks from education and learning science. For example, Bloom \etal~outline three domains of learning: \textit{cognitive} (intellectual skills related to knowledge and problem solving), \textit{affective} (attitudes, values, and interests), and \textit{psychomotor} (ability to physically accomplish tasks) \cite{bloom1956taxonomy}. Krathwohl further revised the taxonomy in the cognitive domain to identify four dimensions of knowledge and six dimensions of cognitive processes. The four dimensions of knowledge are \textit{factual} (\eg~terminology), \textit{conceptual} (\eg~categories, generalizations, and models), \textit{procedural} (\eg~procedures and algorithms), and \textit{metacognitive} (\eg~strategies and self-knowledge). The six dimensions of cognitive process are \textit{remember}, \textit{understand}, \textit{apply}, \textit{analyze}, \textit{evaluate}, and \textit{create} \cite{krathwohl2002revision}. 


Two major themes emerge from these bodies of work. First, the notion of skill is closely associated with the ability to accomplish a task. Second, skills can be expressed as the possession of different types of knowledge. In the context of this paper, we can thus define chart authoring skills as the ability to create specific charts with given datasets. Such skills entail the acquisition of knowledge such as conceptual knowledge (\eg~the meaning of mark and axis), procedural knowledge (\eg~how to compose a chart using a specific tool), and metacognitive knowledge (\eg~how to choose a different construction strategy if the current one fails).

This definition, however, is too high-level to be useful in identifying and discussing the skills needed for chart authoring.
Enumerating different types of knowledge required for an authoring task is practically impossible. Kusterer calls this the ``infinite regress problem'': even for the most mundane skill, it is impossible to write down everything that one needs to know \cite{kusterer1978workplace}. 
On the other hand, while task is a well-studied topic in the data visualization literature, most efforts focus on tasks involved in \textit{using} visualizations for data analysis, with few attempts to formulate chart authoring tasks. To address this gap, we decided to explore characterizing chart authoring tasks by identifying subtasks at multiple levels of granularity across different stages in the authoring process as described in the next subsection.


\subsection{Analysis of Tasks and the Indispensability of Tools}
\label{sec:tool_importance}

In our attempt to characterize chart authoring skills, we tried to develop a taxonomy of tasks. Inspired by Brehmer and Munzner's approach \cite{brehmer2013typology}, we focused on the \textit{multi-faceted} and \textit{multi-level} nature of chart authoring tasks. A task is multi-faceted in the sense that it can be described along three dimensions: \textit{why} the task is performed, \textit{how} the task is performed, and \textit{what} inputs the task takes and outputs it produces. A task can also be described at multiple levels of abstraction, \eg~high-level (consume) to mid-level (search) to low-level (query) for the \textit{why} dimension \cite{brehmer2013typology}. 

Similar to Brehmer and Munzner, we started by formulating abstract authoring tasks that were independent of the properties and designs of specific tools. We observed that many authoring tasks could be grouped into a small number of recurring high-level approaches. For example, some tools require authors to compose charts from primitives, while others allow authors to instantiate predefined templates or examples. These observations led to the identification of recurring authoring paradigms that later became one of the dimensions in the framework. In addition, many common tasks such as data transformation and visual mapping, as identified in the InfoVis Reference Model \cite{card_readings_1999}, depend heavily on the characteristics of input data and output chart. For example, the specific data transformations required depend on both the input data format and the output chart structure. Instead of enumerating all such individual tasks in the framework, we chose to incorporate input data and output chart as framework dimensions, allowing the framework to capture the underlying factors determining skill requirements in a more concise way.

Once we began describing chart authoring tasks at the lower levels, however, we realized that it is difficult to do so without referring to the properties, designs, and capabilities of the authoring tool to be used. This observation challenged common wisdom and approach in skill characterization and visualization task abstraction. For instance, none of the theories from psychology and social sciences we reviewed (Section \ref{sec:skill_theory}) tried to incorporate tools as a critical part of the formulation of skills or competence. Existing frameworks of visualization tasks also argue that the specific features of a tool may not be a top priority. When theorizing about the role of interaction in data visualization, for example, Yi \etal observed that ``for different representation techniques, different interaction techniques are used to perform a similar task or achieve a similar goal,'' and thus it made sense to focus on ``what users achieve by using the interaction techniques rather than how the techniques provided by Infovis systems work'' \cite{yi_toward_2007}. Brehmer and Munzner also argued that a task typology that abstracts away the tool features provides ``a consistent lexicon for description that supports making precise comparisons of tasks between different visualization tools and across application domains'' \cite{brehmer2013typology}. The \textit{how} part of their typology thus focuses on abstract methods associated with interaction techniques to encode data, and to manipulate or introduce visual elements.

This line of approach makes sense when the goal is to achieve a tool-agnostic and domain-independent understanding of visualization tasks. However, for our purpose of characterizing user skills in chart authoring, it is not meaningful to focus on abstract tasks alone without considering tools. For example, when authoring a multi-series line graph (Scenario 2) using two different tools (Altair and Power BI) with the same input dataset, the high-level task (compose) and the lower-level tasks (e.g., import data, specify encodings) can all be the same, but different tools entail different skill requirements. Some skills are transferable across tools (e.g., knowledge of the concept of encoding), whereas others are tool-dependent (e.g., data binding via programming vs. GUI). 
In addition, the same tool can also be used in different ways to create the same chart, involving different tasks and entailing different skills. For example, \rev{users can use Charticulator~\cite{ren2018charticulator} to create a ribbon chart either from a predefined template or from scratch.} The former expects authors to prepare the input data in a template-compatible format while the latter requires them to \textit{compose} a chart in addition to preparing the data in a tool-compatible format. \rev{These observations suggested that characterizing chart authoring skills requires considering not only the tasks being performed but also the tools through which those tasks are accomplished.}


\subsection{\rev{Analysis of Tools and Synthesizing CAPED}}
\label{sec:synthesis}
Given the diversity and variety of chart authoring tools, and the fact that existing authoring tools keep evolving and new tools emerging, it is not feasible to enumerate all the possible tools and their features. We thus searched for higher-level concepts that remain stable across tools, and integrate the accounts on tasks and tools in a unified framework.  

We initially considered the following dimensions for characterizing the various aspects of chart authoring tools: system model (how charts are internally represented and updated in a tool), 
user interface (the means through which users operate the tool and receive feedback on their actions), 
modality (the input/output channel between the user and the authoring tool), 
device (the hardware device with which users perform chart authoring), and rendering (how a tool renders a chart, e.g., raster or vector graphics).
We then tried to apply these dimensions to analyze the authoring experiences in various tools, which made us realize several limitations in focusing on these dimensions. 

While system models are important in understanding how a tool works, relevant details are not always available, especially in commercial tools. 
Furthermore, it is usually more important for users to understand the tool's vocabulary or conceptual building blocks than its internal representations.
For example, users do not have to understand VizQL to construct visualizations in Tableau; they just need to have conceptual knowledge of visual encodings that specify the mappings between the data attributes and the visual channels. 
On the other hand, while UI, modality, device, and rendering can capture important tool properties from a human-computer interaction perspective, they are insufficient in capturing tool properties specific to data visualization. For example, we cannot differentiate the skills required by programming approaches to chart authoring such as D3 and Vega-Lite because they share the UI (code editor), modality (mouse and keyboard), device (desktop), and rendering (SVG). Throughout this process, two dimensions emerged as more appropriate than the aforementioned dimensions: the conceptual abstraction exposed by a tool (e.g., monolithic prototype, parameterized template, building block construction, or data-bound vector-graphics) and the means through which users express intent (e.g., natural language, widget manipulation, table editing, drawing, or programming languages).

Finally, we synthesized these insights into a unified framework organized through the lens of tasks and tools. The resulting framework was iteratively refined through the analysis of representative authoring scenarios, comparisons across chart authoring tools, and discussions of skill transfer between authoring environments.

\section{The CAPED Framework: Authoring Skills through the Lens of Tasks and Tools}
\label{sec:framework}
\rem{In this section, we introduce} \rev{The CAPED framework consists of five dimensions that determine the knowledge and abilities required in a chart authoring scenario: output \textbf{C}hart, chart \textbf{A}bstraction, \textbf{P}aradigm, \textbf{E}xpression, and input \textbf{D}ata. The acronym \textit{CAPED} evokes the idea of being ``equipped'' with the skills and expertise needed for chart authoring, much like a hero wearing a cape. The five dimensions are described below:}

\begin{itemize}[itemsep=.5mm]
    \item \textit{Input data} (task - what): the dataset to be visualized, which may be clean, tidy, tool-compatible, or template-compatible. As explained earlier, these different types of datasets imply different levels of skills. For example, preparing a tool-compatible dataset requires the conceptual knowledge of the format assumed by a chosen authoring tool as well as the ability to transform a dataset into the required format. 
    \item \textit{Output chart} (task - what): the target \rev{chart} design to be created. Authoring a particular chart usually requires the ability to describe, read, and understand the chart. 
    \item \textit{Paradigm} (task - how): the high-level approach to author a chart. Example paradigms include composition, instantiation, delegation, and repurposing. The paradigm is closely associated with the authoring tool. For example, D3 users tend to compose charts. However, the same tool may support multiple paradigms, and the same output chart may be created using different paradigms in the same tool. 
    \item \textit{Abstraction} (tool - what): the conceptual framework of what constitutes a \rev{chart}. Example abstractions include parameterized templates, building block construction, data-bound vector-graphics, and monolithic prototype. 
    \item \textit{Expression} (tool - how): the medium through which users articulate their intents. Expressions are closely related to abstractions. Example expressions include table editing, programming in DSL, UI widget manipulation, and natural language utterances. 
\end{itemize}   


\subsection{Task-Related Dimensions: Input, Output, Paradigm}

The characterization of chart authoring tasks focuses on the \textit{what} and \textit{how} aspects. The \textit{why} dimension is straightforward with a single objective at the top level: to author a \rev{chart}. At the next level, we identify two tasks that correspond to the underlying reasons (or contexts) for authoring: understand data during the exploratory data analysis and present the data or insights for the communication purposes.
We \rem{next} \rev{thus} elaborate on the {what} and {how} dimensions.

\subsubsection{What: Input Data and Output Chart}
We identify two types of input for a \rev{chart} authoring task at the top level for the \textit{what} dimension: (1) dataset and (2) a \rev{chart} template or example required by some tools (\eg~Chartreuse \cite{cui_mixed-initiative_2022} and Mystique \cite{chen_mystique_2023}). The output of the authoring task is a functional \rev{chart} that encodes the input data. 
At a lower level, the input dataset may be characterized as:
\begin{itemize}
    \item \textit{Clean} data: dataset without errors, missing values, and redundancies, and thus ready to be used as is.
    \item \textit{Tidy} data: clean data organized into a consistent structure \cite{wickham_tidy_2014}.
    \item \textit{Tool-compatible} data: dataset formatted to be compatible with the target tool. For example, Tableau \cite{noauthor_tableau_nodate} expects the data table to be in the long format \cite{wickham_tidy_2014}, so that attributes can be dynamically mapped to visual channels. 
    \item \textit{Output-compatible} data: many tools not only require the dataset to be in a specific format (e.g., long or wide), but also assume that the dataset conforms to a particular schema that is compatible with the output. 
    For example, Mystique~\cite{chen_mystique_2023} analyzes an example SVG chart to infer the required number of categorical and quantitative columns, and performs a compatibility check on the user's data. 
\end{itemize}
  
These dataset categories are useful when dealing with authoring tools that require different skills in data transformation. For instance, tools like Falx \cite{wang_falx_2021} and Data Formulator \cite{wang_data_2024} require the input data to be tidy, alleviating the burden of performing complex data transformations (\eg~pivoting), while many authoring tools require the input data to be formatted in a tool-compatible manner \cite{satyanarayan_critical_2019}. 






\subsubsection{How: Authoring Paradigm}
For the \textit{how} dimension, we identify the following four tasks, each describing an authoring paradigm characterized by the amount of available scaffolding that embodies the desired chart output.


\begin{itemize}
\item \textit{Compose:}~authoring a chart by composing a set of predefined primitives or building blocks from scratch. Examples include composing a radial network diagram with hierarchical edge bundling using D3's cluster layout and radial line generator, or building a multi-series line chart by dragging and dropping data attributes to the construction shelf in Power BI (Scenario 2). In both cases, tool-compatible datasets are required. 


\item\textit{Instantiate:} authoring a chart by infusing new data into a template or chart example supported by a tool. Examples include creating a waterfall chart using Excel's built-in template (Scenario 1), or creating a diverging stacked bar chart using Mystique \cite{chen_mystique_2023} by providing an SVG example. In both of these cases, tool-compatible datasets are required. 


\item\textit{Delegate:} authoring a chart by delegating the task to a human or a software agent. Examples include generating a parallel coordinates plot by prompting ChatGPT (Scenario 3), or creating a stacked area chart through Tableau's Show Me recommender \cite{mackinlay_show_2007}. Tidy data is required as input in the former case, while tool-compatible data is required in the latter case. 

\item\textit{Repurpose:} modifying the components (e.g., mark, layout, encoding) of an existing chart to author the desired chart output. Examples include creating a dumbbell chart by modifying a scatter plot in Excel (Scenario 1). In this case, tool-compatible input data is still required. 


\end{itemize}

We want to emphasize that these four tasks are not meant to be exhaustive. They instead represent common high-level authoring approaches observed in research and practice. 
Furthermore, accomplishing an authoring goal may require a series of these four authoring tasks. In Scenario 1, for example, Alex needs to first \textit{instantiate} a scatter plot, and then \textit{repurpose} it to create a dumbbell chart. 

These tasks can be complemented by low-level descriptions of the actions and operations involved. For example, to compose a multi-series line graph in Scenario 2, Kori needs to first \textit{transform} data to prepare the input data, then to \textit{import} the data into the authoring tool (Power BI), and finally to \textit{configure} the construction shelf. Each of these actions may be decomposed into even lower-level operations. For example, configuring the construction shelf involves \textit{operations} such as \textit{dragging and dropping} columns to the corresponding shelves.

\subsection{Tool-Related Dimensions: Abstraction and Expression}

\rev{Chart authoring tools can be characterized along two key dimensions: Abstraction and Expression.}

\begin{table*}[h!]
\centering
\setlength{\tabcolsep}{6pt}       
\renewcommand{\arraystretch}{1.3} 
\arrayrulecolor{lightgray}
\begin{tabular}{| p{1.3cm} | p{2.4cm} | p{3cm} | p{1.7cm} | p{3.4cm} | p{3.3cm} |}
    \hline
    \rowcolor[gray]{.95} \textbf{Tool} & 
    \textbf{Input Data} & 
    \textbf{Output Chart} &
    \textbf{Paradigm} & \textbf{Abstraction} & \textbf{Expression} \\
    \Xhline{2\arrayrulewidth}
    \multirow{2}{*}{Excel} & template-compatible data & waterfall chart & instantiate  & monolithic prototype &  widget manipulation + \newline table editing \\
    \cline{2-6} & 
    template-compatible data & dumbbell chart & instantiate \( \rightarrow \) \newline repurpose & monolithic prototype \( \rightarrow \) data-bound vector-graphics  &  widget manipulation + \newline table editing\\
    \Xhline{2\arrayrulewidth}
    \multirow{2}{*}{ChatGPT} & clean data & grouped bar chart & delegate \( \rightarrow \) compose & monolithic prototype \( \rightarrow \) building block construction & natural language + DSL \\
    \cline{2-6} & 
    clean data & parallel coordinates plot & delegate  & monolithic prototype    & natural language \\
    \hline
\end{tabular}
\caption{Two examples of authoring different charts using the same tool, Excel \rev{(Scenario 1)} and ChatGPT \rev{(Scenario 3)}.} 
\label{tbl:excel}
\end{table*}

\begin{table*}[h!]
\centering
\setlength{\tabcolsep}{6pt}       
\renewcommand{\arraystretch}{1.3} 
\arrayrulecolor{lightgray}
\begin{tabular}{| p{2cm} | p{2.6cm} | p{2cm} | p{1.4cm} | p{3.5cm} | p{3.5cm} |}
    \hline \rowcolor[gray]{.95}
    \textbf{Output Chart} &
    \textbf{Input Data} & 
    \textbf{Tool} &
    \textbf{Paradigm} & \textbf{Abstraction} & \textbf{Expression} \\
    \Xhline{2\arrayrulewidth}
    \multirow{2}{*}{\makecell[l]{multi-series\\line chart}} & tool-compatible data & Altair & compose  & building block construction & declarative DSL \\
    \cline{2-6} & 
    tool-compatible data & PowerBI & instantiate & parameterized template  & widget manipulation\\
    \Xhline{2\arrayrulewidth}
    \multirow{2}{*}{\makecell[l]{dumbbell\\chart}} & clean data & Data\newline Formulator & instantiate +  delegate  & parameterized template & natural language, widget manipulation, DSL \\
    \cline{2-6} & 
    clean data & Falx & delegate + instantiate & building block construction & structured object demonstration, widget manipulation \\
    \hline
\end{tabular}
\caption{Two examples of authoring the same chart, multi-series line chart \rev{(Scenario 2)} and dumbbell chart, using different tools.} 
\label{tbl:same_chart_diff_tool}
\end{table*}

\subsubsection{Abstraction} 

\rev{Chart} abstractions are the conceptualizations of \rev{charts} exposed to the user through a tool's interface. 
We identified the following types of \rev{chart} abstractions:
\begin{itemize}
    \item \textit{Monolithic Prototype}: in this abstraction, a \rev{chart} is conceptualized using chart typology\rem{ or an example image}. For example, a chart created using Excel's template \rem{is usually treated as a monolithic object in that it} is associated with a chart type name and a representative icon. \rev{ Customization is primarily limited to styling and reference elements like axis and legend.} 
    \item \textit{Parameterized Template}: many tools employ a top-down approach to \rev{chart} authoring, where users start with a chart \rem{type} template and fill in the data attributes to be mapped to various visual channels. For example, the Ivy system \cite{mcnutt_integrated_2021} allows users to select \rem{one} from a collection of parameterized templates, and then \rev{set the parameters by manipulating widgets or editing a textual specification.} 
    \item \textit{Building Block Construction}: in this approach, the abstraction is the product of compiling a specification, which is often declarative. This kind of abstraction is found in libraries based on the Grammar of Graphics \cite{wilkinson_grammar_2005} such as Vega-Lite. \rev{It also includes authoring tools built on top of such specifications such as Lyra~\cite{satyanarayan2014lyra}.} \rem{There are two main characteristics of this abstraction. First, each tool identifies a set of building blocks that capture various aspects of a \rev{chart} (e.g., mark, encoding),} \rev{Users create data visualizations in a bottom-up approach by assembling chart building blocks (e.g., mark, encoding). Modifying the chart involves changing the specification at the level of the building blocks, either through a programming API or a GUI.}
    \item \textit{Data-Bound Vector-Graphics}: \rev{a chart can also be viewed as a vector graphic in which certain visual properties are determined by data. The incorporation of data binding can be achieved either through a programming API (\eg D3.js) or a GUI (\eg Data Illustrator \cite{liu_data_2018}).}
\end{itemize}

\rem{A chart created by an AI chatbot may also be viewed as a monolithic object: the user's prompt may include an example image or a desired chart type name. However, current AI tools typically rely on building block specification as their underlying representation. Therefore, when automatic generation fails, users may have to work at the building block specification abstraction.}

It is important to note that the same type of abstraction can be used in different authoring paradigms. For example, both the instantiate and delegate paradigms can treat an output chart as a monolithic prototype. Conversely, the same authoring paradigm may be realized based on different abstractions. To compose a chart, for example, one can choose a GoG library (building blocks specifications) or D3.js (data-bound vector-graphics); Charticulator allows users to create \rev{charts} through either building block construction or parametrized templates. 



\subsubsection{Expression} 
Expression is the medium through which users articulate their intents when authoring a \rev{chart}. While abstraction is about \textit{what} kind of conceptual descriptions underlie a chart, expression is about \textit{how} that conceptual description is articulated by users in the tool. Expression implicitly captures the corresponding user interface, modality, and device, but places a stronger emphasis on user action and the target object of the action. We identify the following types of expressions: 

\begin{itemize}
    \item \textit{Table editing}: in tools like Excel, \rev{users create charts} through static templates. Users edit \rev{data} in a spreadsheet interface so that the data format and schema are compatible with the chosen template. 
    \item \textit{Drawing}: users can ``sketch'' partial or complete data visualizations with a digital pen or a mouse to convey the desired design. The drawing can be casual and low-fidelity (\eg Data Ink \cite{xia2018dataink}), or high-fidelity with precise control over the details (\eg using professional tools like Adobe Illustrator).
    \item \textit{Gesturing}: users express their intents through mouse, touch (\eg on touch screens) or spatial gestures (\eg in VR environments).
    \item \textit{Structured object demonstration}: users edit \rev{chart} components \rev{by example to illustrate and extrapolate to the whole chart} (\eg Falx \cite{wang_falx_2021}, StructGraphics \cite{tsandilas_structgraphics_2020})  
    \item \textit{Widget manipulation}: users manipulate widgets to control visual properties or \rev{shelf configuration (\eg Tableau \cite{noauthor_tableau_nodate})}.
    \item \textit{Domain-Specific Language (DSL)}: users employ declarative (e.g., Vega-Lite \cite{satyanarayan_vega-lite_2016}, ggplot2 \cite{wickham_ggplot2_2009}) and imperative (\eg Mascot~\cite{liu_manipulable_2025}, PICCL~\cite{haoyan_shi_piccl_2025}) programming languages.
    \item \textit{Natural language}: users convey their intents in natural language (\eg conversational agent), using either as typed text or as \rev{speech-to-text} input.
    \item \textit{Example chart}: in some tools, users can convey the target chart by providing an example in a tool-compatible format. 
    For example, Mystique~\cite{chen_mystique_2023} expects an example \rev{chart} in the SVG format, while Charticulator~\cite{ren2018charticulator} requires the \rev{chart} template to be a tool-compatible JSON file, containing information necessary to render a chart with a template-compatible input data. 
\end{itemize}

The same expression may be used with different \rev{chart} abstractions. For example, widgets can be used on both monolithic prototypes and parameterized templates. Conversely, the same abstraction may be articulated using different expressions: building block specification may be done using widget manipulation, natural language, or DSL.

\section{Applying the Framework}

\subsection{Using the Framework to Describe}
\label{sec:describe}

In this section, we demonstrate how to apply the CAPED framework to describe skills required in different authoring situations, and use the framework to describe skills as knowledge. \rev{This section refers to the scenarios explained in Section~\ref{sec:scenarios}. To follow along with the chart examples, please visit our supplemental website: \url{https://caped-framework.github.io/}.}

\subsubsection{Describing Authoring Scenarios}









\noindent\textbf{Authoring different charts using the same tool.} 
Table \ref{tbl:excel} summarizes different authoring activities using the same tool, including those described in \href{https://caped-framework.github.io/#scenario-01}{Scenario 1 (Excel)} and \href{https://caped-framework.github.io/#scenario-03}{Scenario 3 (ChatGPT)}. We use \( \rightarrow \) to denote the transition between different stages of authoring that involve different paradigms, and + to denote the combined use of multiple abstractions or expressions during a stage. For example, 
in \href{https://caped-framework.github.io/#scenario-03}{Scenario 3}, Jesse delegates the authoring of two different charts to an AI assistant. Depending on the outcome, different abstractions and expressions are involved. When AI generates the desired output, the \rev{chart} abstraction can stay at the level of monolithic prototype; however, once something goes wrong, Jesse needs to dive into the abstraction level of building block construction to inspect the code produced by AI in a domain-specific language. 


\bpstart{Authoring the same chart using different tools} In \href{https://caped-framework.github.io/#scenario-02}{Scenario 2}, Kori can author the same multi-series line graph using Altair or Power BI, as shown in Table~\ref{tbl:same_chart_diff_tool}. The demonstrative differences in skill are the paradigm (compose vs. instantiate),  abstraction (buidling block construction vs. parameterized template) and expression
(a declarative DSL vs. widget manipulation). 
We hypothesize that skill transfer between Altair and Power BI requires relatively low effort compared to other skill leaps between two tools. However, skill transfer is not symmetric: widgets are more familiar to users than DSLs. 

In other cases, two tools might appear similar at first glance (e.g., Data Formulator \cite{wang_data_2024} and Falx \cite{wang_falx_2021}), but their abstractions require users to approach authoring with different mental models of how the input data realizes an output chart. Table \ref{tbl:same_chart_diff_tool} demonstrates how users would use Data Formulator and Falx, two tools designed to alleviate the burden of data transformation in \rev{chart} authoring, to create a dumbbell chart. They both operate under the same authoring paradigms: instantiate (picking a chart type in Data Formulator and choosing an inferred chart design in Falx) and delegate (Data Formulator and Falx send queries to LLMs and program synthesis algorithms to transform data, respectively). Authors experience a notable difference in the abstraction between these two tools. With Falx, they express the desired \rev{chart} using a miniature version of the target \rev{chart} - through visualization-by-demonstration they abstractly construct the chart. In Data Formulator, instead of constructing, authors pick a chart type and then fills in the shelf configuration with existing or desired data columns (called \textit{data concepts}). When necessary, authors refine descriptions for data concepts using natural language or table editing, and the system generates data transformations to realize the desired data concepts.
Since Falx does not expose data concepts as an abstraction, authors have no access to the inferred data transformation logic. In contrast, Data Formulator presents the generated data transformation code along with the transformed table, supporting authors to identify and correct potential mistakes from the delegated data transformation task. Interpreting data transformation code is a skill needed when using Data Formulator, but not Falx.



\bpstart{Authoring the same chart using the same tool} Even when we use the same tool to author the same chart, it is possible to have different authoring processes that involve different skills. 
Consider \href{https://caped-framework.github.io/#line-chart-example}{Line Chart Example} to create a faceted line chart using ggplot2 (Table~\ref{tbl:same_chart_same_tool}). The  faceted line chart shows stock prices (y-axis) during the same time periods (x-axis) for three companies: IBM, Microsoft (MSFT), and Amazon (AMZN). ggplot2 incorporates methods to manipulate input data and output charts, supporting alternative design approaches to authoring the same chart with two different input data schemas. Given data in the long format with ``date'', ``company'', and ``price'' columns, authors can employ the \texttt{facet\_grid} operator to create faceted sub line charts with the column ``company''; if the input data is in the wide format with ``date'', ``MSFT'', ``AMZN'', and ``IBM'' columns, authors can apply different configurations to the same base chart using \texttt{lappy} to produce a repeated line chart for each company. 

\begin{table*}[ht]
\centering
\setlength{\tabcolsep}{4pt}       
\renewcommand{\arraystretch}{1.3} 
\arrayrulecolor{lightgray}
\begin{tabular}{|p{2cm} | p{1.9cm} | p{3.2cm} | p{1.6cm} | p{3.5cm} | p{3.7cm} |}
    \hline \rowcolor[gray]{.95}
    \textbf{Output Chart} & \textbf{Tool} &
    \textbf{Input Data} & 
    \textbf{Paradigm} & \textbf{Abstraction} & \textbf{Expression} \\
    \Xhline{2\arrayrulewidth}
    \multirow{2}{*}{\makecell[l]{faceted\\line chart}} & \multirow{2}{*}{ggplot2} & long-format data & compose  & building block construction & declarative DSL \\
    \cline{3-6} &  &
    wide-format data & compose & building block construction  & declarative + imperative DSL\\
    \hline
    \Xhline{2\arrayrulewidth}
    \multirow{2}{*}{\makecell[l]{ribbon\\chart}} & \multirow{2}{*}{Charticulator} & tool-compatible data & compose  & building block construction & drawing, widget manipulation \\
    \cline{3-6} &  &
    template-compatible data & instantiate & parameterized template  &  widget manipulation\\
    \hline
\end{tabular}
\caption{Two examples of authoring the same chart (faceted line chart and ribbon chart) using the same tool (ggplot2 and Charticulator) but following different authoring approaches due to the differences in the input data.}
\label{tbl:same_chart_same_tool}
\end{table*}

\begin{table*}[h!]
\centering
\setlength{\tabcolsep}{4pt}       
\renewcommand{\arraystretch}{1.3} 
\begin{tabular}{|p{2cm} | p{1.9cm} | p{3.2cm} | p{1.6cm} | p{3.5cm} | p{3.7cm} |}
    \hline \rowcolor[gray]{.95}
    \textbf{Output Chart} &
    \textbf{Tool} &
    \textbf{Input Data} & 
    
    \textbf{Paradigm} & \textbf{Abstraction} & \textbf{Expression} \\
    \Xhline{2\arrayrulewidth}
     Sankey\newline diagram & StructGraphics & not required \( \rightarrow \) template-compatible data & compose \( \rightarrow \) instantiate & data-bound graphics \( \rightarrow \) parameterized template  & drawing \( \rightarrow \) table editing \\
     \hline
     tower chart & Mystique & template-compatible data & instantiate & parameterized template & widget manipulation \\
     \hline
\end{tabular}
\caption{Additional examples of unconventional \rev{chart} authoring approaches given unique output charts (Sankey diagram and tower chart).}
\label{tbl:more_examples}
\end{table*}

\begin{table*}[t]
\centering
\renewcommand{\arraystretch}{1.5}

\begin{tabularx}{\textwidth}{|p{2.1cm}|X|X|X|X|}
\hline

\rowcolor{gray!10}
 &
\multicolumn{1}{c|}{\textbf{Compose}} &
\multicolumn{1}{c|}{\textbf{Instantiate}} &
\multicolumn{1}{c|}{\textbf{Delegate}} &
\multicolumn{1}{c|}{\textbf{Repurpose}} \\
\hline

\rowcolor{orange!10}
&
\multicolumn{4}{l|}{%
For the data: Understand the data requirements of the tool and the visualization to be created.%
} \\

\cline{2-5}
\rowcolor{orange!10}
\multirow[t]{-2}{2.1cm}{%
  \textbf{Conceptual Knowledge} (examples)
}
&
Understand the meaning, application, and relationship of primitives.
&
Understand the meaning of the desired template and its availability.
&
Assess the delegate tool's capability, results, and output quality.
&
Understand the underlying frameworks for the source and the target. \\
\hline
\rowcolor{blue!10}
&
\multicolumn{4}{l|}{%
For the data: Prepare the data in the required input format.%
} \\
\cline{2-5}
\rowcolor{blue!10}
\multirow[t]{-2}{2.1cm}{%
  \textbf{Procedural Knowledge} (examples)
}
&
Conduct the required steps to construct a chart with primitives.
&
Identify and instantiate the appropriate template.
&
Specify the chart with valid semantics and syntax for the given language.
&
Convert the source chart using the target data. \\
\hline

\rowcolor{green!10}
& \multicolumn{4}{l|}{%
Know the capabilities and limitations of the tool and choose an appropriate tool.%
} \\

\rowcolor{green!10}
& \multicolumn{4}{l|}{%
\raisebox{0.6ex}{Understand the strengths and weaknesses of different authoring strategies and choose an appropriate one.}%
} \\

\rowcolor{green!10}
\multirow[t]{-3}{2.1cm}{%
  \textbf{Meta-cognitive Knowledge} (examples)
}

& \multicolumn{4}{l|}{%
\raisebox{1.4ex}{%
Decide to switch to a different tool for a better outcome or adapt to different strategies for error recovery.%
}%
} \\[-0.2ex]
\hline

\end{tabularx}

\caption{Example knowledge for each of the four authoring paradigms, focusing on the tools and including considerations for the data.}
\label{tbl:knowledge}
\end{table*}

In \href{https://caped-framework.github.io/#ribbon-chart-example}{Ribbon Chart Example}, Charticulator provides two ways to create a ribbon chart (Table~\ref{tbl:same_chart_same_tool}): by constructing from scratch (\textit{paradigm}: compose), or by using a ribbon chart template (\textit{paradigm}: instantiate). 
In this example, a ribbon chart shows mobile operating system market share from 2009 to 2016.
The input data includes ``operating system'', ''year'', and ``share'' columns. 
For an author to compose this ribbon chart, they start by encoding a rectangle glyph's height by ``share'' and the x-axis of the chart by ``year''.
Then they encode the ``operating sytem'' to the color and apply a \textsf{Stack Y} to the glyphs and order the stack by ``share.''
Finally to create a ribbon chart, the author applies a ``Band'' link, connected by the ``platform'' data attribute.
The input data needs to be tool-compatible with Charticulator to compose the ribbon chart from scratch.
However, in the other use case -- to instantiate the chart from a template -- the input data needs to be exactly compatible with the given template. 
Consequently, the abstractions and expressions involved are different for creating the same chart with the same tool.



\bpstart{Different abstractions using the same tool} Finally, we consider a scenario which further demonstrates how different dimensions in the CAPED framework can be combined in unconventional \rev{chart} authoring approaches. 
First, in StructGraphics \cite{tsandilas_structgraphics_2020}, the authoring process consists of two stages (Table~\ref{tbl:more_examples}): 1) users start by composing graphics on a canvas without the prerequisite of an input dataset, the graphical attributes of the composed graphics form a data table structure that can be assigned to data, like a template; 2) the author then instantiates this bespoke template they authored in Structgraphics by applying compatible data to it. In the first stage, the \rev{chart} abstraction is data-bound vector-graphics, accomplished through a drawing interface; in the second stage, the abstraction is a template. 

Similar to StructGraphics, Mystique \cite{chen_mystique_2023} derives a data table structured from vector graphics, sharing the instantiation paradigm and template-compatible input data with the StructGraphics approach (Table~\ref{tbl:more_examples}). However, Mystique does not require users to have the skills to draw desired \rev{chart}s in a canvas. Instead, it uses a mixed-initiative approach to analyze the graphical structure in an existing SVG chart. As a result, users need to be able to specify the mappings between data attributes and visual properties through a wizard interface composed of interactive widgets. 


\subsubsection{Describing Skills as Knowledge}
One way to capture the skills required for chart authoring is to identify the conceptual, procedural, and meta-cognitive knowledge associated with the tasks. We acknowledge that it is impossible to enumerate all possible knowledge for all the lower-level tasks because of the ``infinite regress problem'' \cite{kusterer1978workplace}. However, it is still beneficial to provide concrete examples of these different types of knowledge to ground our discussion. 
Table~\ref{tbl:knowledge} includes some example knowledge for each of the four authoring paradigms, focusing on the tools. We formulate knowledge in the form of verb + noun phrases, following the convention used in education and learning science \cite{bloom1956taxonomy,krathwohl2002revision}. 

\textbf{Conceptual knowledge} involves understanding the meaning of \rev{chart} elements and the frameworks that underlie chart creation. The ability to understand the kind of data required by a chosen tool (e.g., tidy, tool-compatible), and the ability to accurately interpret the \rev{chart} to be created are essential for all four authoring paradigms. To be able to \textit{compose} a \rev{chart} using a chosen tool, users must understand the meaning of the primitives provided by the tool and the relationships between them. They should also be able to give examples for each primitive. To \textit{instantiate} a \rev{chart}, users must be able to know whether the desired template to create the \rev{chart} is provided by the tool, and understand what a desired template means. To \textit{delegate} the creation task, users should be able to reliably estimate the capability of the delegate and the quality of the output, as well as to correctly interpret the results generated by the tool. Finally, to \textit{repurpose} an existing chart, users must understand the underlying mechanisms in the tool for the creation of both the source chart and the target chart.  

\textbf{Procedural knowledge} refers to the ability to carry out the concrete steps needed to author \rev{charts} using a tool's interface. A common type of procedural knowledge shared across the authoring paradigms is the ability to prepare data in the required input format. In the \textit{compose} paradigm, this involves executing the construction steps by combining the primitives through the tool interface. With \textit{instantiate}, procedural knowledge is required to identify and apply templates, and then populate them with data through the user interface. In the \textit{delegate} paradigm, an example of procedural knowledge is to specify the desired \rev{chart} using the language (e.g., natural language, demonstration) accepted by the tool. For \textit{repurpose}, users need to be able to re-specify the target design or make adjustments through the interface. 

\textbf{Meta-cognitive knowledge} encompasses awareness of the strengths and limitations of different authoring options, and the ability to make strategic choices for a specific authoring goal. For example, users need to know the capabilities and limitations of the tool and choose an appropriate tool. Once a tool is chosen, users should understand the pros and cons of different construction strategies and choose an appropriate one. When the authoring process stalls, meta-cognitive knowledge entails making a decision to switch to a different tool for a better outcome or adapt to different strategies for error recovery.

\subsection{Identifying New Research Opportunities}
\label{sec:identify}



To minimize the skills and time required for \rev{chart} authoring, many researchers and technologists envision an authoring process, where users simply describe their goal (sometimes with an example image) and AI generates the desired chart in a single step. Such a process focuses on a specific combination of three dimensions in the CAPED framework: delegation (\caped{paradigm}) + monolithic prototype (\caped{abstraction}) + natural language (\caped{expression}). 
However, due to practical constraints (e.g., incomplete or ambiguous user prompts, limited model performance, high task complexity), AI may fail to generate the correct chart automatically. \rev{More broadly, the rapid emergence of diverse chart authoring tools, including AI-powered systems, raises new questions about how different users should be supported, both through intelligent authoring systems and through visualization education. The CAPED framework provides a conceptual tool for reasoning about different aspects of authoring skills and identifying} 
 opportunities 
\rev{for supporting} users with varying backgrounds and abilities. 
Below, \rev{we discuss} three examples.


\bpstart{Opportunity 1}
For users who can interpret AI-generated results, having skills related to compose (\caped{paradigm}) and building block construction or data-bound vector-graphics (\caped{abstraction}), we argue that it is important to support seamless transitions between different \caped{paradigms}, \caped{abstractions}, and \caped{expressions}. Natural language can be an effective type of \caped{expression} to ask AI to generate initial \rev{chart} specification or code. However, it is not always the most appropriate \caped{expression} for effectively communicating intent (e.g., describing bespoke \caped{outputs}), or for tasks such as refining mark properties (e.g., increasing bar width) and chart styles (e.g., changing color schemes). Other forms of \caped{expressions}, such as drawing and widget manipulation, may be more effective in certain situations. To transition from natural language to these expressions, the corresponding \caped{paradigms} and \caped{abstractions} may need to change as well. For example, to support easy refinement of mark properties, AI-generated D3 code (where the underlying \caped{abstraction} is data-driven graphics) needs to be adapted into building block construction, so that users can directly manipulate the properties through a dynamically generated GUI. 




\bpstart{Opportunity 2}
\rev{The CAPED framework offers an opportunity in \rev{chart} pedagogy by helping educators tailor learning materials to different tools and  learners' prior skills. Rather than treating chart authoring as a fixed sequence of steps, the framework can distinguish the skills that learners need to acquire from those already supported by a tool.
For example, a business professional moving from Excel to Power BI may already be familiar with fitting data into parameterized templates (\caped{abstraction}) and manipulating widgets (\caped{expression}). For this type of learner, instructional materials can focus less on chart types or data transformation, but more on the compose \caped{paradigm} and visual encoding concepts needed to create more expressive charts.}
 \rev{The same applies for introducing users to new tools such as Mystique \cite{chen_mystique_2023} and Data Formulator \cite{wang_data_2024,wang2025data}. In these cases, learning materials 
need not emphasize manual data transformation or chart specification when these tasks are automated, but instead should help users develop the remaining skills required to effectively guide, interpret, and refine chart designs. More broadly, CAPED provides educators with a vocabulary for reasoning about learning objectives across authoring tools, enabling curricula to emphasize enduring authoring competencies rather than tool-specific procedures.
}
\bpstart{Opportunity 3}
In the third example, consider a graphic designer who does not know how to code---building block construction (\caped{abstraction}) and Domain-Specific Language (\caped{expression}).
Like the business professional in the second example, they are familiar with parameterized template (\caped{abstraction}) and widget manipulation (\caped{expression}) but lack skills to perform the necessary data transformations to prepare their data into the expected input data format (\caped{input data}). However, they are likely to be able to describe the target chart (\caped{output chart}) in natural language instruction (\caped{expression}) and more skilled in conveying the target chart with free-form drawing (e.g., sketching using a pen, manipulating shapes using a mouse). In addition, they are comfortable working at the level of vector graphic primitives like shapes and paths (\caped{abstraction}). Existing tools like DataInk \cite{xia2018dataink} and StructGraphics \cite{tsandilas_structgraphics_2020} introduce interaction techniques to support users with these skills, but they do not help with the necessary tasks (i.e., data transformation) they lack skills in. Unlike the second example where AI can generate a parameterized template from drawing, in this case, the designers may benefit more from a tighter integration between AI and drawing interfaces that allow them to create bespoke \rev{charts}.

\section{Discussion and \rev{Future Work}}
\label{sec:discussion}

CAPED is the first theoretical framework to characterize user skills in static chart authoring through the lens of tasks and tools. \rev{In this section, we reflect on the framework’s characteristics, scope, and limitations, as well as the opportunities for future research}.

The five dimensions in our framework are not strictly orthogonal: categories across dimensions may exhibit interdependencies and correlations. For example, natural language instruction as an \caped{expression} is often associated with the delegate  \caped{paradigm} and the monolithic prototype \caped{abstraction}. However, as discussed in \Cref{sec:framework}, associations between specific dimensions' categories are common but not deterministic, and no dimension can be reduced to another. We emphasize that such relationships between the dimensions do not constitute a limitation. Many influential frameworks in HCI and visualization similarly employ dimensions that are conceptually distinct yet partially overlapping. For example, in the Cognitive Dimensions of Notations framework \cite{green1989cognitive,blackwell2001cognitive}, we can identify dimensions that interact with one another (e.g., abstraction and visibility, consistency and viscosity). The observed interdependencies reflect the intrinsic coupling among the tool and task considerations in \rev{chart} authoring. 



We note that the CAPED framework is neither exhaustive nor definitive. As we mentioned in Section \ref{sec:scope}, we chose not to consider interaction and animation to have a manageable scope in this first attempt to characterize authoring skills. Authoring \rev{data} visualization with interaction support would require different skills depending on the complexity of interaction and its underlying visualization. For example, some modern visualization libraries (e.g., Highcharts, Plotly) provide tooltips by default, and thus do not require additional skills to include tooltips through a simple mouseover interaction. Popular business intelligent systems with visualization support (e.g., Microsoft Power BI, Tableau) offer a simple way to include basic interactivity, such as filtering, drill-down, and brushing \& linking. Building highly interactive systems to support sophisticated data exploration would require programming and beyond, such as user interface and interaction design. \rev{Future work could extend the CAPED framework to encompass animation and interaction. Doing so would likely require revising the \caped{output chart} dimension or introducing additional dimensions to capture non-static charts, as well as expandinging the \caped{expression} dimension to include programming languages as another form of expression.}

\rev{Furthermore, we acknowledge that the current framework is primarily derived from the analysis of chart authoring tools within the information visualization domain.
As a result, the five identified dimensions may not fully capture the abstractions employed in scientific or spatial visualization systems, such as the Visualization Toolkit (VTK) \cite{schroeder_visualizing_2000} and Paraview \cite{ahrens2005paraview}.
Data-flow pipelines, computational operators, and spatial data transformations important to these systems substantially differ from the chart-centric authoring systems analyzed in this work.
Therefore, extending and validating this framework to encompass scientific and spatial visualization remains an open area for future exploration.}


The framework identifies a set of core dimensions for \rev{chart} authoring skills, and can serve as a foundation for future work to investigate how to measure or quantify skill levels along one or more of these dimensions. For example, researchers could investigate how to assess prompting proficiency in chart authoring systems, or to systematically evaluate candidates' expertise with a visualization library. 
In addition, researchers could develop standardized assessments similar to literacy or proficiency tests. Such assessments would have several practical applications. They could enable more principled recruitment of participants for user studies, ensuring clearer grouping of participants. They could also provide clearer criteria for job descriptions, as well as for candidate screening and skill evaluation. 
Overall, CAPED opens new opportunities for systematically understanding, assessing, and operationalizing \rev{chart} authoring skills.

\acknowledgments{Bongshin Lee was supported in part by the Institute of Information and Communications Technology Planning and Evaluation (IITP) Grant funded by the Korean Government (MSIT), Artificial Intelligence Graduate School Program, Yonsei University, under Grant RS-2020-II201361. Zhicheng Liu was supported in part by NSF grant IIS-2239130.}

\bibliographystyle{abbrv-doi-hyperref}

\bibliography{authoringskills}

\end{document}